\documentstyle[12pt,epsfig]{iopart}
\newcommand{\us}{\mbox{$\mu$s}}
\newcommand{\C}{\mbox{$^{12}$C}}
\newcommand{\Cd}{\mbox{$^{13}$C}}

\newcommand{\N}{\mbox{$^{12}$N}}
\newcommand{\Nd}{\mbox{$^{13}$N}}

\newcommand{\numu}{\mbox{$\nu_{\mu}$}}
\newcommand{\numub}{\mbox{$\bar{\nu}_{\mu}$}}
\newcommand{\nue}{\mbox{$\nu_{e}$}}

\newcommand{\nueb}{\mbox{$\bar{\nu}_{e}$}}
\newcommand{\nubx}{\mbox{$\bar{\nu}_{x}$}}
\newcommand{\ep}{\mbox{e$^{+}$}}
\newcommand{\el}{\mbox{e$^{-}$}}
\newcommand{\pos}{\mbox{e$^{+}$}}
\newcommand{\mup}{\mbox{$\mu^{+}$}}
\newcommand{\pip}{\mbox{$\pi^{+}$}}
\newcommand{\mupdecay}{\mbox{\mup\ $\rightarrow\:$ \pos $\!$ + \nue\ + \numub}}
\newcommand{\pipmup}{\mbox{\pip $\rightarrow\:$ \mup + \numu}}

\newcommand{\CCd}{\mbox{\Cd\,(\,\nue\,,\,\el\,)\,\Nd }}
\newcommand{\CCprot}{\mbox{p\,(\,\nueb\,,\,\ep\,)\,n }}
\newcommand{\excl}{\mbox{\C\,(\,\nue\,,\,\el\,)\,\N$_{\rm g.s.}$}}
\newcommand{\CCexc}{\mbox{\C\,(\,\nue\,,\,\el\,)\,\N$^{*}$}}
\newcommand{\pn}{\mbox{p\,(\,n,$\gamma$\,)}}

\newcommand{\nuebx}{\mbox{\nueb $\rightarrow\,$\nubx }}
\newcommand{\numunue}{\mbox{\numu $\rightarrow\,$\nue }}
\newcommand{\numubnueb}{\mbox{\numub $\rightarrow\,$\nueb }}
\newcommand{\NCL}{\mbox{90\%\,CL}}
\newcommand{\NNCL}{\mbox{99\%\,CL}}
\newcommand{\Dm}{\mbox{$\Delta m^2$}}
\newcommand{\sit}{\mbox{$\sin ^2(2\Theta )$}}

\begin{document}
% Journal identifier can be put here if required, e.g.
%\jl{14}

\title[Compatibility Analysis of LSND and KARMEN]
      {Compatibility Analysis of the LSND Evidence and the
       KARMEN Exclusion for \numubnueb\ Oscillations}

\author{Klaus Eitel\footnote{Address as of October 1st, 1999:
Forschungszentrum Karlsruhe, P.O. Box 3640, 
D-76021 Karlsruhe, Germany}}

\address{Los Alamos National Laboratory, Los Alamos, NM 87545, USA}

\begin{abstract}
A combined statistical analysis of the experimental results of the
LSND and KARMEN \numubnueb\ oscillation search is presented. 
LSND has evidence for neutrino oscillations that is
not confirmed by the KARMEN experiment. However, there is a region in the
(\sit ,\Dm) parameter space where the results of both experiments are
statistically compatible. This joint analysis is based on likelihood 
functions for both data sets. A frequentist approach creating Monte Carlo
samples analogous to the experimental outcome is applied to deduce correct 
confidence limits. Different schemes of combination can be chosen to provide 
correct coverage which lead to slightly different confidence regions 
in (\sit ,\Dm).
\end{abstract}

\pacs{06.20.Dk, 14.60.Pq} % measurement and error theory
			  % neutrino mass and mixing

\section{Introduction}\label{intro}
% ----------------------------------
Over the last years, the controversial results of the two experiments
LSND (Liquid Scintillator Neutrino Detector at LANSCE, Los Alamos, USA) and
KARMEN (KArlsruhe Rutherford Medium Energy Neutrino experiment at ISIS,
Rutherford, UK) both searching for neutrino oscillations \numubnueb\ have
led to intense discussions. The two experiments are similar as they use
\numub\ beams from the \pip -\mup\ decay at rest (DAR) chain \pipmup\ 
followed by \mupdecay\ with energies up to 52\,MeV. Furthermore,
both experiments are looking for \nueb\ from \numubnueb\ oscillations via the
reaction \CCprot\ providing a spatially correlated delayed coincidence
signature of a prompt \pos\ and a subsequent neutron capture signal.

However, the detection techniques are significantly different:
LSND uses a homogenous detector volume of mineral oil with a low 
concentration of scintillator viewed by 1220 phototubes, thereby giving 
excellent particle identification by detecting
a directional Cherenkov cone as well as isotropic scintillation
light with a characteristic pattern of hit photomultipliers 
\cite{Ldet}. KARMEN is a segmented liquid scintillation calorimeter with
excellent time and energy resolution exploiting the distinct time structure
of the ISIS neutrino source \cite{Kdet}. Thus, at KARMEN a \nueb\ excess
from \numubnueb\ would be identified by requiring its time
distribution to follow the 2.2\,\us\ slope from the parent \mup\ decay.

In two data sets taken during the periods 1993--95 and 1996--98 LSND
has observed a clear beam--on minus beam--off excess of events with
\nueb\ signature, i.e. (\pos ,n) sequences. These have been interpreted as 
evidence for \numubnueb\ 
oscillations (see \cite{atha} for the first data set). The analysis of the
sum of these data sets, although slightly different in their spectral shape,
results in corresponding favored areas of the mixing parameter \sit\ and 
\Dm\ \cite{rexllwi}.

KARMEN has found no excess events above the expected background.
For all events, potential \nueb\ signal and measured background, the energy, 
time and spatial distributions for both the prompt and delayed events are 
precisely known. Using this spectral information also leads to no hint for
oscillations. Therefore, KARMEN cannot confirm the LSND result.
Furthermore, \NCL\ exclusion limits are deduced cutting into
the LSND evidence region in the (\sit,\Dm) parameter space for \numubnueb\
oscillations \cite{karmenos},\cite{rmwin}.

The statistical analysis of the data has become a showcase of how to determine 
statistical significance and upper limits. KARMEN with no apparent \numubnueb\ 
signal and very low background has the problem of treating a result in a
low statistics regime near the physical boundary $\sit = 0$. In
LSND, the maximum likelihood analysis of the data clearly
indicates an oscillation signal. A problem arises when determining a
region of correct confidence, i.e. statistical significance, in the 
(\sit ,\Dm) plane having a likelihood function in two parameters, which shows 
a pathological behavior, namely an oscillatory dependence in \Dm\ 
with numerous local maxima.

In 1998, the discussion was intensified 
by a paper of Feldman and Cousins \cite{cous}, who described a method
of dealing with the problems described above. Their approach
to extract upper limits in the case of a small number of measured events
is itself highly controversial, though recently adapted by the Particle Data 
Group~\cite{pdg}.

This report describes the individual evaluation of both data sets with
maximum likelihood methods. The statistical interpretation of the
likelihood functions and confidence regions is based on a frequentist approach 
and follows closely the analysis suggested by Feldman and Cousins.
The main purpose of such an approach is to determine correct regions of
confidence in (\sit,\Dm)\footnote{In a slightly different manner, a similar
analysis was performed much earlier, for the G\"osgen reactor
experiment \cite{zacek}.}. A correct coverage is defined in terms of
frequency, i.e. fraction of occurrence for future experiments.
Probability or confidence in this context does not mean ``degree of belief''
as defined in a Bayesian statistics.
For a detailed introduction into Bayesian and frequentist approaches we
refer to~\cite{dagos}.

Although the central statements of LSND and KARMEN are contradicting there
can be a region in the (\sit,\Dm) parameter space where the results 
are compatible. Combining the two experiments is done in different ways of 
constructing statistical distributions, pointing out that there is no unique 
way of determining regions of specific confidence. However, as we will see,
the regions of compatibility in (\sit ,\Dm) are very similar.

The method described below is a complete analysis of the
two experiments. However, the actual result is preliminary for various 
reasons: A new analysis of the LSND data is under way with a new
reconstruction algorithm. In addition, flux calculations and efficiencies for 
1996-1998 used here are still preliminary. The KARMEN2 data used for this 
analysis is taken from February 1997 to February 1999 and represents about 50\%
of the envisaged total accumulated neutrino flux for the upgraded experiment. 
It is therefore an intermediate data set updated by the ongoing experiment.

A statistical analysis combining two experimental results which apparently
disagree is a delicate and controversial approach. It is not
the task nor the purpose of this analysis to overcome this disagreement.
However, assuming that there is no serious systematical error in either of
the experiments and the interpretation of their results with respect to
oscillations \numubnueb , the question of statistical compatibility of the 
individual results is well justified and should be addressed quantitatively. 
This is the objective of the analysis presented in this paper.

\section{KARMEN2 data evaluation}\label{K-data}
% ---------------------------------------------

In 1996, the KARMEN experiment underwent a substantial upgrade.
An additional veto counter with 300\,m$^2$ surface
surrounding the central detector on all sides
was the main improvement. This veto counter reduces
cosmic induced background for the \numubnueb\ search by a factor of 40,
which consists of energetic neutrons produced in the iron of the blockhouse
by deep inelastic scattering of cosmic muons.
With the new configuration and increased neutron detection efficiency,
KARMEN is running as KARMEN2 since February 1997. Starting as a simple
counting experiment \cite{nu98}, the evaluation method was changed last summer
to a more sophisticated maximum likelihood analysis of the data,
making use of detailed event information in energy, time and spatial position.

\subsection{The data set}

\begin{figure}[hbt]
  \centerline{\epsfig{figure=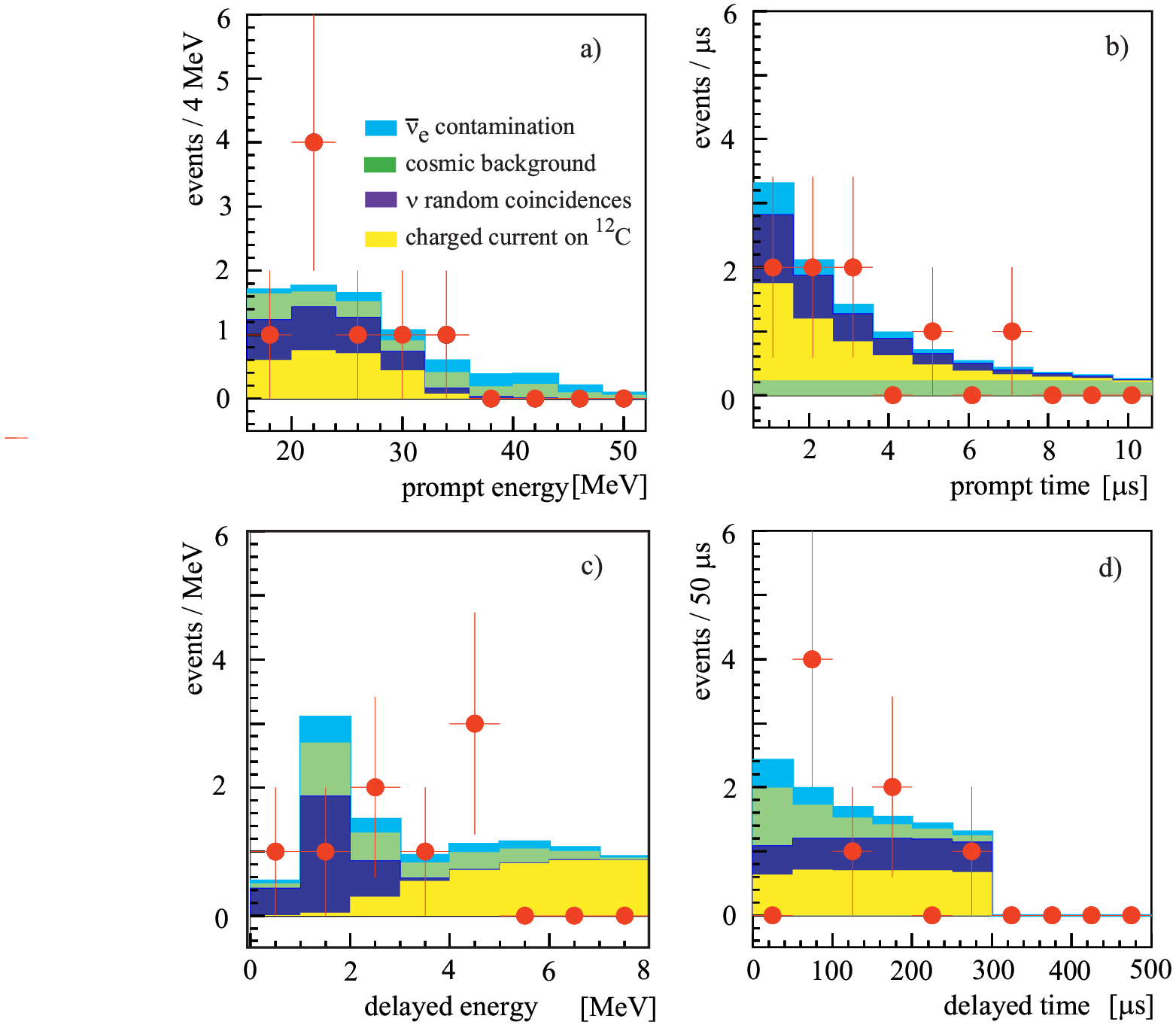,width=13.0cm}}
  \caption{Remaining sequences after applying all cuts. Superimposed
	to the energy (a,c) and time (b,d) of the initial and the delayed
	event are the expected background contributions.}
  \label{viererplot}
  \end{figure}

The data collected through February 1999 correspond to 4670\,C accumulated
proton charge on the ISIS target. Veto cuts for all veto components up to 
24\,\us\ before a potential oscillation event were applied and 
a spatial coincidence between the initial \pos\ and the neutron capture of
1.3\,m$^3$ was required. Figure~\ref{viererplot} summarizes the remaining 8 
event sequences in the appropriate energy and time windows.
The background components are also given with their distributions.
All components except the intrinsic \nueb\ contamination are measured
online in different time and energy windows (see table~\ref{bgsum}).
\begin{table}[htb]
\caption{Background contributions to the \numubnueb\ search
	and their determination methods.}
\hspace{0.2cm}
\label{bgsum}
\begin{indented}
\item[]
\begin{tabular}{l|c|l}
background & expectation & determination \\ \hline
\raisebox{0pt}[10pt][7pt]{\excl\ sequences} 
  & \raisebox{0pt}[10pt][7pt]{$2.6\pm0.3$} 
  & \raisebox{0pt}[10pt][7pt]{measurement in diff. E/t-window} \\
\raisebox{0pt}[10pt][7pt]{induced by cosmic $\mu$} 
  & \raisebox{0pt}[10pt][7pt]{$2.3\pm0.3$} 
  & \raisebox{0pt}[10pt][7pt]{pre-beam measurement} \\
\raisebox{0pt}[10pt][7pt]{$\nu$-induced accidental coinc.} 
  & \raisebox{0pt}[10pt][7pt]{$1.9\pm0.1$} 
  & \raisebox{0pt}[10pt][7pt]{measurement in diff. E/t-window} \\
\raisebox{0pt}[10pt][7pt]{ISIS source contamination} 
  & \raisebox{0pt}[10pt][7pt]{$1.1\pm0.1$} 
  & \raisebox{0pt}[10pt][7pt]{MC simulation} \\ \hline
\raisebox{0pt}[10pt][7pt]{total background} 
  & \raisebox{0pt}[10pt][7pt]{$7.8\pm0.5$} & \\
\end{tabular}
\end{indented}
\end{table}
In total, the background expectation amounts to $7.8\pm0.5$ events. Therefore, 
the 8 extracted sequences show no hint for an oscillation signal.
For oscillations with full mixing and large \Dm , i.e. 
($\sit=1,\Dm \ge 100$\,eV$^2$), a signal of $1605\pm176$ coincidences were
expected. Taking the results from section~\ref{feldman} for the LSND signal
in the region $\Dm < 2$\,eV$^2$,
one would have expected an oscillation signal of about 2 to 6 \numubnueb\ 
events added to the background within the data set. In order
to extract more information from the 8 events about any potentially small
oscillation signal a detailed maximum likelihood analysis was performed.

\subsection{Data analysis}

This likelihood function analyses 5 event parameters: the
energy $E_p$ and $E_d$, the prompt time $t_p$ and the delayed coincidence 
$\Delta t=t_d-t_p$ as well as the spatial correlation
$\Delta \vec{x}=\vec{x}_d-\vec{x}_p$.
The likelihood is calculated varying the oscillation 
signal $r_{osc}$ as well as the background components relative to the
overall data sample:
$r_{CC}$ for charged current events \excl , $r_{cos}$ for cosmic background,
$r_{ran}$ for random coincidences with a $\nu$--induced prompt event and
$r_{con}$ for the intrinsic \nueb\ contamination. With the condition
$\sum_{j=1}^5 r_j=1$ and $\rho=(r_{osc},r_{CC},r_{cos},r_{ran},r_{con})$ the 
likelihood function for the $M=8$ events can be written as
\begin{eqnarray} \nonumber
  L(\rho) &=& \prod_{k=1}^M \{ \sum_{j=1}^5 r_j \cdot f_{j1}(E_p^k) \cdot
	 f_{j2}(E_d^k) \cdot f_{j3}(t_p^k) \cdot f_{j4}(\Delta t^k) 
	 \cdot f_{j5}(\Delta \vec{x}^k) \} \\ 
	 & \times & \prod_{j=2}^5 P(r_j|r_j^{expected})
 \label{kar-lhd}
\end{eqnarray}
The density functions $f_{ji}$ contain the spectral information of all 
components, and as the positron energy spectrum depends on \Dm , the
dependence of $L$ on \Dm\ enters via the density function $f_{11}$.
The parameter \sit\ is determined by the ratio of oscillation events
$N_{osc}=M\cdot r_{osc}$ divided by the expected number of events for maximal 
mixing $N_{exp}(\Dm,\sit=1)$: $\sit = N_{osc}/N_{exp}$.
The second line in (\ref{kar-lhd}) is the combined Poisson probability 
$\prod P$ for the background contributions $r_j$ calculated with the 
expectation values $r_j^{expected}$.

As mentioned before, all background components but the 
simulated intrinsic contamination
are measured online with their spectral information $f_{ji}$ and the
expectation values $r_j^{expected}$. Since the event sample is still very 
small, the different backgrounds are summarized into one component with
fixed relative contributions:
$f_{bi}= \sum_{j=2}^5 r_j^{expected} \cdot f_{ji}$, and
the combined Poisson probability in (\ref{kar-lhd}) reduces to
$P(r_b|r_b^{expected})$ with $r_b^{expected}=7.8/8$. Hence, the likelihood 
function effectively has two free parameters, $r_{osc}$ or \sit\ and \Dm .

For technical reasons, it is more convenient to optimize the logarithmic 
likelihood function $lnL$. Taking the 8 events of KARMEN2 so far, this 
function has its maximum at a value of $\sit<0$. This can be explained by the 
measured energy spectrum of the prompt events (figure~\ref{viererplot}a).
Since there are all events below 36\,MeV but $1.0\pm 0.1$ background events
expected above $E_p=36$\,MeV, the best fit consists in increasing slightly the
overall background contribution. Since an oscillation signal for large \Dm\
has a higher contribution for $E_p>36$\,MeV, a negative oscillation signal
can then compensate the background to account for no events in that energy
region.

\begin{figure}[hbt]
  \centerline{\epsfig{figure=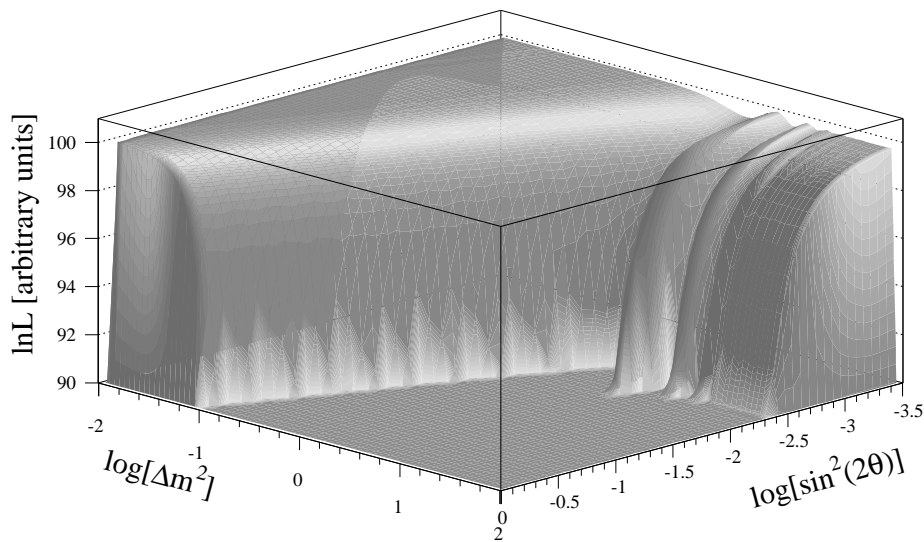,width=12.0cm}}
  \caption{Logarithmic likelihood function $lnL(\sit,\Dm)$ for the 8 events
	of KARMEN2. The maximum in the physically allowed region $\sit \ge 0$
	is set to a value of 100, the minimum of this plot is set to 90.}
  \label{k-lhdfunction}
  \end{figure}

Figure~\ref{k-lhdfunction} shows $lnL$ where the
maximum in the physically allowed range $\sit\ge 0$ has been renormalized
to a value of $lnL(\sit_m=0,\Dm)=100$. Note the sharp fall with increasing 
\sit .
From the likelihood function it is obvious that there is no oscillation signal
in the data. The task at this point is how to extract upper limits in
(\sit,\Dm) for a given confidence level.

The method of extracting correct confidence regions in (\sit,\Dm) is based
on a frequentist approach and will be discussed in detail in 
section~\ref{feldman}: The basic idea is to create a large number of event 
samples analogous to the experiment. These samples are created by Monte Carlo
using the full event information for the likelihood procedure.
The samples also contain oscillation events based on a hypothesis $H$ with
$(\sit_H,\Dm_H)$. A statistic is constructed by comparing each sample's 
maximum of $lnL$ with the value $lnL(\sit_H,\Dm_H)$ from which confidence 
regions are extracted.

\section{LSND data evaluation}\label{L-data}
% ------------------------------------------

The data analysed in this context have been reduced by requiring
the following criteria: They are so-called 'electron-like' events surviving a
$\chi_L$ cut, they have energies of $20\le E\le 60$\,MeV and their 
reconstructed distance to the tank photomultiplier surface is $d\ge 35$\,cm.
In the following, all data from 1993 through 1998 are analysed in one
data set, where the flux calculations, efficiencies, and background treatment 
of the 1996-1998 data are preliminary extrapolations of the older data subset.
The information about a delayed event is compressed into a likelihood
ratio R. If within one millisecond after the initial event another event 
is recorded at distance $\Delta r\le 250$\,cm, the ratio of likelihood R in
energy (PMT hits), time and distance of being a correlated \pn\ over an
accidental coincidence is calculated, otherwise $R=0$.
Details of the event reconstruction and the definition of R can be found in
\cite{Ldet} and \cite{long}.

\subsection{Event samples}

\begin{figure}[hbt]
  \centerline{\epsfig{figure=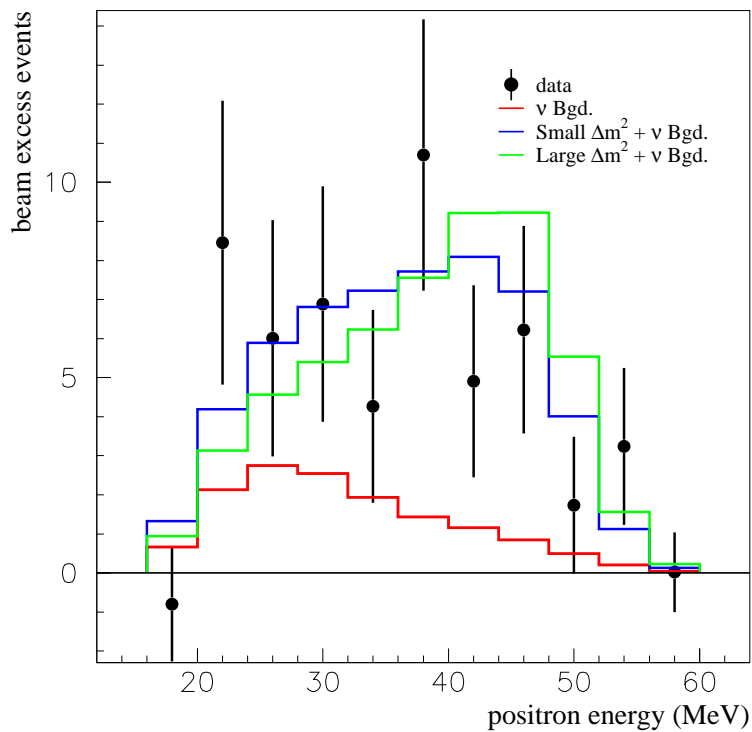,width=10.0cm}}
  \caption{LSND 1993-1998 beam excess applying a cut of $R\ge 30$.
	The excess after subtracting beam-related background is
	$39.5\pm8.8$ events. The probability for being a statistical
	fluctuation is less than $10^{-6}$.}
  \label{gold}
  \end{figure}

Requiring a high likelihood ratio $R$ selects (\pos,n) correlated events with
low uncorrelated background. Applying a cut of $R\ge 30$ reduces the data
to 70 events with $20\le E\le 60$MeV. After subtracting beam unrelated 
($17.7\pm1.0$) and beam related ($12.8\pm1.7$) background, a net 
excess of $39.5\pm8.8$ events remains. Its energy distribution is shown
in figure~\ref{gold} and clearly demonstrates the significance of the
excess.
This sample is referred to by LSND as ``gold-plated'' events.
Taken the excess as oscillation signal, the signal to background ratio
for this sample is better than 1.

To determine the oscillation parameters \sit\ and \Dm , an event
sample with no cut in R is used, leading to higher efficiency for the 
oscillation channel, but naturally increasing the background.
  \begin{figure}
  \centerline{\epsfig{figure=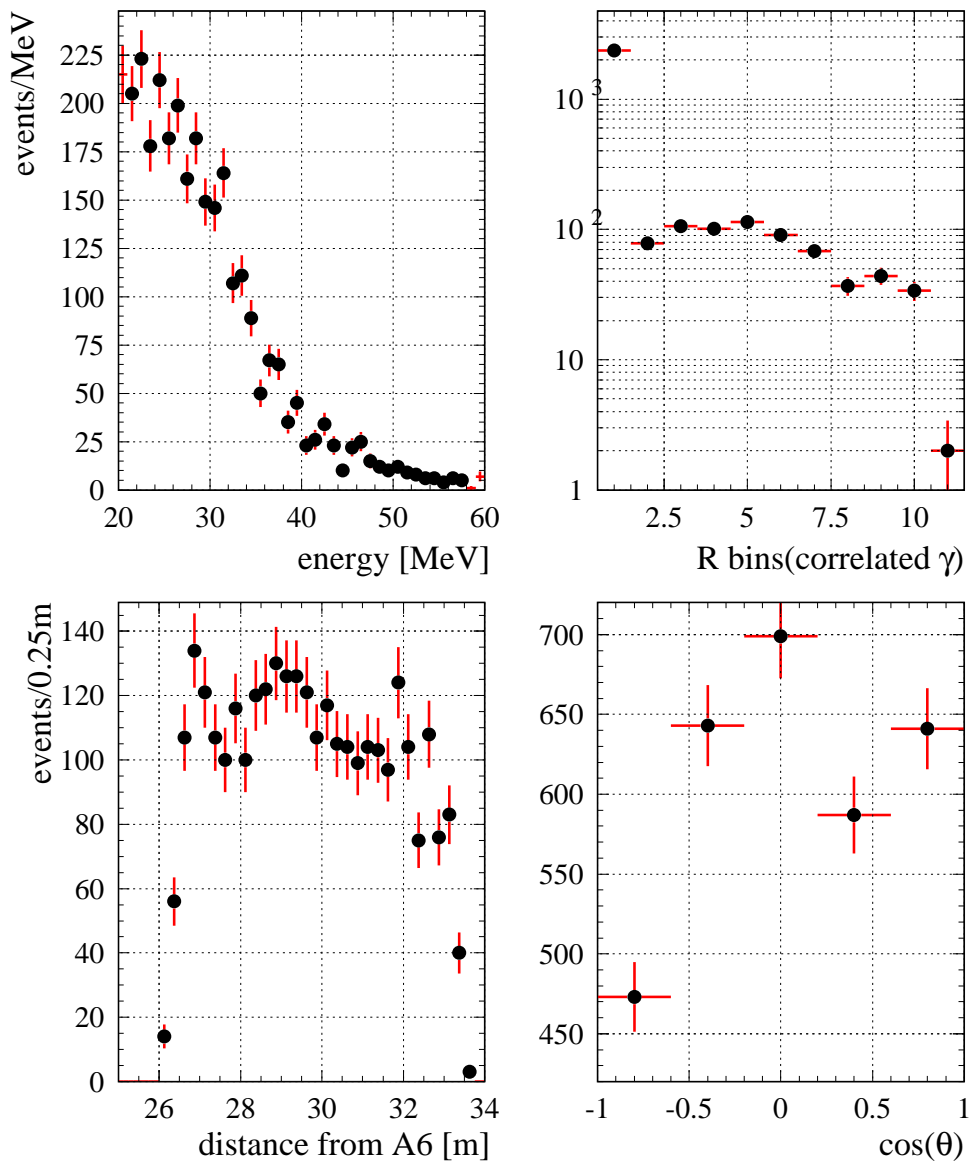,width=13.0cm}}
  \caption{LSND 1993-1998 data set with loose cuts containing 3049 events
	including a total beam-off background expectation of $1467\pm38$ 
	events. The bin number N in R corresponds to an upper bound of
	$R=10^{N/3-1}$.}
  \label{data}
  \end{figure}
Figure~\ref{data} shows this event sample comprising 3049 beam-on events. 
Four variables are used to categorize
the events: Electron energy,
likelihood ratio R, spatial distribution in the detector expressed in distance 
L to the LANSCE neutrino source A6 and the angle $\cos \theta$ between 
the direction of the incident neutrino and the reconstructed electron path.
Note that these are projections of a 4-dim space of correlated parameters
for each event.

The evaluation method uses these 4 correlated parameters 
to extract the oscillation signal, i.e. \sit\ and \Dm , from other
background sources by calculating the overall likelihood function $L$ for all
3049 events. There are 2 ways of setting up this function described below.

\subsection{Maximum likelihood analysis of bins}\label{bin_subsec}

Up to the evaluation of the 1997 data, the likelihood function was defined as
the product of the Poisson probability of all bins in the 4-dimensional
space made up by the parameters $(E,R,L,\cos \theta)$. Although this
definition will not be used in the further context, the analysis of the
1993--95 data in \cite{long} was based on this definition of the likelihood 
function. It is therefore described here
for completeness and to allow a comparison to the results of the new
likelihood method given in section~\ref{evt_subsec}.

The bin--based likelihood function was set up as
\begin{equation}
	 L(n_1,n_2,\dots| \Dm , \sit ) = \prod_{i=1}^N \frac{1}{n_i! }
		\nu_i^{n_i} e^{-\nu_i}
 \label{bin-lhd}
\end{equation}
with $N=198000$ bins, $n_i$ the number of beam-on events in bin $i$ and 
$\nu_i$ the expected event number:
\begin{equation}
	\nu_i = \nu_i(BUB) + \nu_i(BRB) + \nu_i(\Dm ,\sit) \quad .
\end{equation}
BUB indicates beam unrelated background measured in the beam off time
window with high precision, BRB stands for beam related background
made up of various components: \excl\ (making up 50.5\% of the total BRB),
\CCexc\ (33.6\%), \nue -\el\ scattering (7\%), \CCd\ (5.7\%), \numu - and
\numub - induced events on protons and \C\ (1.9\%) and intrinsic \nueb\ 
contamination (1.1\%).
The relative contributions are subject to change due to the 
1996--1998 preliminary evaluation.

The definition of the likelihood function was straight forward, but
the treatment of the errors on the expected background components was somewhat
arbitrary. The best fit resulted in an oscillation signal which, added up to
the expected backgrounds, didn't explain the total beam excess, i.e. there
were events in the sample which were not attributed to either BUB, BRB or
oscillations. In a second step, the likelihood function was calculated
again with a beam related background expectation enhanced (lowered) by
one standard deviation. All three resulting $lnL$-functions were put in .OR.,
which means added with the same relative weight. The problem with this
approach is that although the likelihood analysis clearly indicates an
upward fluctuation of the BRB, the case of lowering the expectation of BRB
is taken with the same weight as $BRB+1\sigma$ therefore artificially
increasing the potential oscillation signal.

The favored region in (\sit,\Dm) was then defined by taking values greater 
than max$(lnL)$-2.3 and max$(lnL)$-4.6 as one would expect to determine \NCL\ 
and \NNCL\ regions in a two dimensional Gaussian likelihood function.
Given the oscillatory behavior of $lnL$ as a function of \Dm\ this gives
obviously only favored regions but no correct coverage in terms of confidence.

\subsection{Event-based likelihood analysis}\label{evt_subsec}

To overcome the problem of background fluctuations, an
alternative construction of the likelihood function based directly on the 
events was introduced \footnote{This method, in addition, has the technical 
advantage of being less CPU time consuming if the number of events investigated
is less than the number of bins in the parameter space.}. 
In the following, all results are based on this new method unless otherwise 
indicated explicitly.

The likelihood function is the product of all $M$ individual event likelihoods
to fit a combination of 4-dim density distributions $f(E,R,L,\cos\theta)$
where the relative strengths $r$ of the contributions are the parameters to be
optimized with the side condition $\sum r_j=1$. In an approximation, all
beam related backgrounds are added up to one contribution
\footnote{Although this is reasonable due to the fact that the accuracy
of all the BRB expectation values is dominated by the same systematic error of
the $\nu$ flux calculation, in the forthcoming new LSND event reconstruction 
and analysis, these components will be considered individually.}.
Treating all beam related backgrounds as one component, the likelihood 
function is defined as
\begin{eqnarray} \fl
  L(r_{osc},r_{brb}) = \prod_{k=1}^M \{ r_{osc}f_{\Dm}(E_k,R_k,L_k,\cos 
  \theta_k) + r_{brb}f_{brb}(E_k,R_k,L_k,\cos \theta_k) \nonumber \\ 
\lo+ (1-r_{osc}-r_{brb})f_{bub}(E_k,R_k,L_k,\cos \theta_k) \} \nonumber \\
 \cdot e^{-\frac{(r_{brb}M - N_{brb})^2}{2\sigma^2_{brb}}}
 \cdot e^{-\frac{(r_{bub}M - N_{bub})^2}{2\sigma^2_{bub}}}
 \label{evt-lhd}
\end{eqnarray}
There are effectively three free parameters: $r_{osc}$ or \sit , \Dm\ 
and $r_{brb}$.
The Gaussian terms account for the background expectation values and their
systematic and statistical uncertainties $BUB=1467\pm38$ and $BRB=1349\pm148$.
If the shape analysis of the $M$ events (first two lines of 
equation~\ref{evt-lhd}) favors values of $r_{brb}$ or
$r_{bub}$ corresponding to BRB or BUB far from the expectation, the
overall likelihood value is reduced by these Gaussian expressions.

The oscillation parameter \sit\ is determined as a function of \Dm\ 
according to
\begin{equation}
	\sit = \frac{r_{osc}\cdot M}{N_{\Dm}(\sit=1)}
\end{equation}
where $N_{\Dm}(\sit=1)$ indicates the number of oscillation events
expected for a given \Dm\ and full mixing $\sit=1$ in the detector, taking 
all resolution functions and cuts into account. The absolute event numbers
corresponding to the maximal value of $L(r_{osc},r_{brb})$ are 73 oscillation 
events, 1495 BRB
and 1481 BUB events for a 3049 total event sample. As already indicated
in the bin-based likelihood analysis, the best fit favors an upward fluctuation
of the beam related background of one standard deviation.

\begin{figure}[hbt]
  \centerline{\epsfig{figure=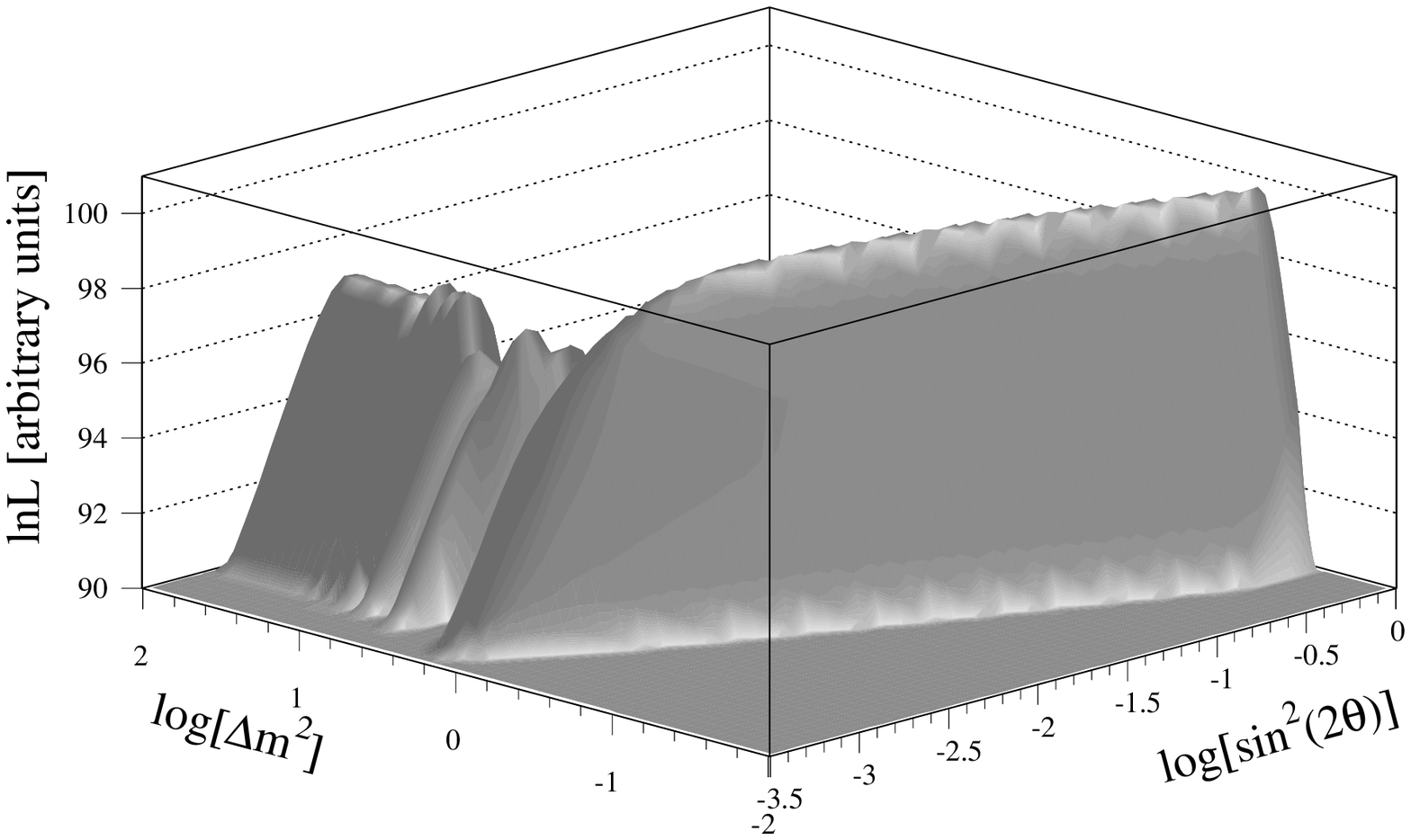,width=13.0cm}}
  \caption{Logarithmic likelihood function $lnL(\sit,\Dm)$ for the LSND
	data 1993-1998 sample containg 3049 events. The maximum in the 
	physically allowed region $\sit \le 1$
	is set to a value of $lnL(\sit_m,\Dm_m)=100$.}
  \label{l-lhdfunction}
  \end{figure}

In a next step, the original likelihood function (\ref{evt-lhd}) is then 
integrated along the axis
of the parameter $r_{brb}$ which is of no further interest. This is a
standard procedure of reducing free parameters of a likelihood function
described in \cite{frode}. The logarithmic
likelihood $lnL$ is therefore a function of the 2 free oscillation
parameters $lnL(\sit,\Dm)$ which is shown in figure~\ref{l-lhdfunction}. 
It reaches its maximum in the physically allowed range of $\sit \le 1$ at 
$(\sit_m=0.93,\Dm_m=0.056eV^2)$ which is set to $lnL(\sit_m,\Dm_m)=100$.
However, the position of this maximum in (\sit,\Dm) is not significant due
to the flatness of the likelihood function along its 'ridge' for small 
values of \Dm . Allowing values of $\sit>1$ in the maximum likelihood analysis,
the maximum of $lnL$ increases by only 0.02 units along a line of
constant values $\sit \cdot (\Dm)^2$ which can be understood by developing
the oscillation probability for small \Dm :
\begin{equation}
	P(\numubnueb) = \sit \cdot \sin^2(1.27\Dm \frac{L}{E})
	\approx \sit \cdot (\Dm)^2 \cdot 1.27^2 \frac{L^2}{E^2}
	\quad .
\end{equation}
In the limit of small \Dm , the energy distribution of oscillation events does
not depend any longer on \Dm . This is underlined by the best fit result of 73 
oscillation events which is stable along the 'ridge' of $lnL$. Note, however,
that the maximum likelihood analysis favors the lowest value of 
\Dm\ possible within the fit, or, in other words, the \nueb\ energy spectrum 
with the lowest mean value possible by oscillations \numubnueb .

\subsection{Checks of the event-based likelihood results}
            \label{evtcheck_subsec}

To test the
dependence of the likelihood function on the expectation values of the
background, the Gaussian terms in (\ref{evt-lhd}) were omitted. The result
of the likelihood analysis is nearly unchanged underlining the stability
of the shape analysis. 

In a second test, only events with $36\le E\le 60$\,MeV
were analysed. The event sample consisted of 476 events with expectations
of $BRB=119\pm13$ and $BUB=274\pm17$ for the background. The maximum of
$lnL$ is then reached at $(\sit=0.69,\Dm=0.063eV^2)$ with 46 oscillation
events, $BRB=148$ and $BUB=281$ using the full likelihood function as
defined in (\ref{evt-lhd}). This result is in good statistical
agreement with the analysis of the larger event sample concerning the
extracted oscillation signal. However, it is
obvious that with the stringent energy cut there is a certain loss of
discrimination power between low and high \Dm\ solutions (see for example
figure~\ref{gold}). This can be seen by comparing the best fit values for
$\Dm=100$\,eV$^2$ . For both energy windows the favored mixing at this \Dm\ is 
$\sit=4.0\cdot 10^{-3}$. The distance in logarithmic likelihood 
units to the maximum is different, however:
$\Delta lnL = lnL(\sit_m,\Dm_m) - lnL(4.0\cdot 10^{-3},100eV^2) = 2.6$ for
$20\le E\le 60$\,MeV compared to $\Delta lnL = 0.25$ for $36\le E\le 60$\,MeV.

Applying the stringent energy cut of $36\le E \le 60$\,MeV and analysing only
the spectral shape of the events (i.e. discarding the Gaussian expectation
terms in equation~\ref{evt-lhd}) leads to a less consistent maximum likelihood
result. Again, there is almost no discrimination between different \Dm\
values: The values of $lnL$ along the best fit values of \sit\ as a function 
of \Dm\ are almost constant. But the oscillation signal is significantly 
reduced from about 46 to some 21 events, with background contributions to
the data sample of $BRB=267$ and $BUB=188$ far from their expectation values.
This may indicate some unexplained distortion of the high energy part of the 
event spectrum but could also be due to the already mentioned naturally
decreased discrimination ability of the maximum likelihood method in a very 
narrow energy window.

\begin{figure}[hbt]
  \centerline{\epsfig{figure=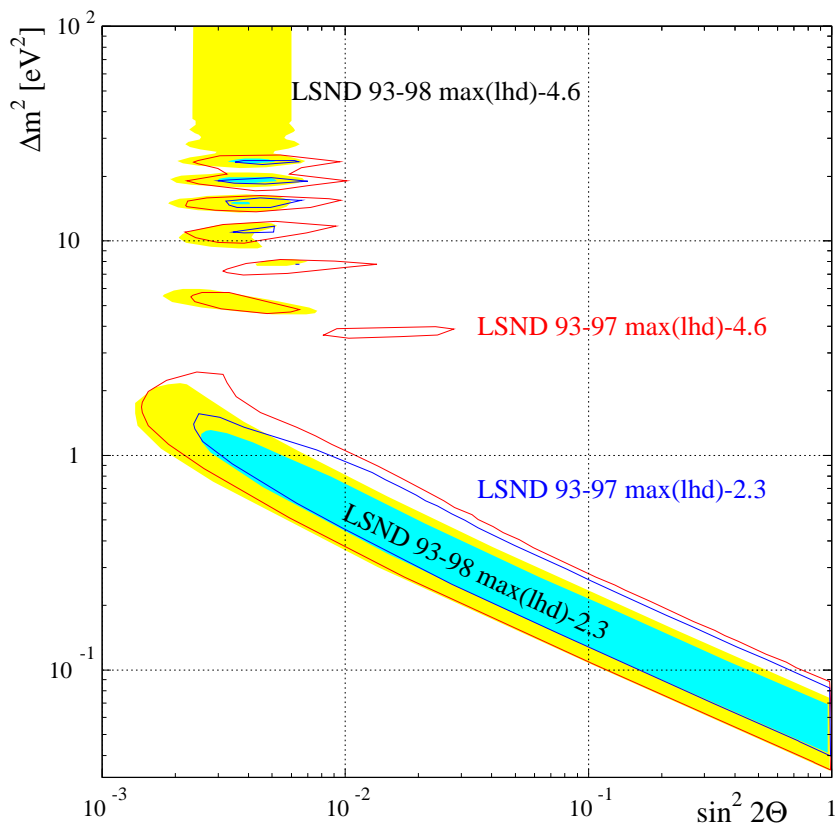,width=11.0cm}}
  \caption{Comparison of the favored regions in (\sit,\Dm) using the
	two different likelihood functions. The 93-97 data are analysed
	with the bin-method (see equation~\ref{bin-lhd}), the 93-98 data set
	with the event-based likelihood function from equation~\ref{evt-lhd}.}
  \label{bin-evt}
  \end{figure}

Another useful check is the comparison of the two likelihood functions
defined in (\ref{bin-lhd}) and (\ref{evt-lhd}). Figure~\ref{bin-evt} shows
the regions in (\sit,\Dm) obtained by taking the contours of both logarithmic
likelihood functions at values of $\Delta lnL=2.3$ and 4.6 units below their 
global maximum.
As explained above, the treatment of the background uncertainty in the
bin based analysis favored larger \sit\ values which is not confirmed if
the background can float freely as it is the case in the event based analysis.
The slightly larger data set of 1993--98 in figure~\ref{bin-evt} has almost 
no influence on the outcome of both methods compared to the older 1993--97 set.
The latest LSND publications (e.g.~\cite{rexllwi}) show these favored
regions for the 1993--98 data set.

\section{Combining both experiments}\label{combi}
% -----------------------------------------------

In this section, we will first construct the correct confidence regions
in the parameter space (\sit,\Dm) of \numubnueb\ oscillations for each 
experiment individually. In the second part (\ref{combi_lnl}),
the logarithmic likelihood functions of both experiments are added
and discussed qualitatively. In section~\ref{methods},
methods of using and combining the obtained experiment statistics
are defined. These will then allow the extraction of combined confidence 
regions. The results of these combination methods are presented
and discussed in part~\ref{combiregio}.

\subsection{Construction of individual confidence regions}\label{feldman}

The basic idea of getting correct confidence regions using the
logarithmic likelihood function $lnL(\sit,\Dm)$ is to create a statistic
based on a frequentist approach. A high number of event
samples is created by Monte Carlo using all experimental information
on the event parameters. Different hypotheses are tested by including in the
generated event samples oscillation events according to the oscillation
parameters (\sit,\Dm). In this section we will describe this method in detail
for the LSND experiment and then show the representative results for both
KARMEN and LSND. The analogous statistical approach for the KARMEN data
can be found in detail in \cite{markus}.

For a preselected \numubnueb\ oscillation hypothesis $H$ with 
oscillation parameters $(\sit_H,\Dm_H)$ the creation of a LSND-like 
event sample is done in two steps.
First, the number of oscillation events, BRB and BUB are thrown on the basis
of the corresponding expectation values \footnote{The overall event number
per sample is fixed to be 3049, the experimental event number.}.
In a second step, for each event, parameters (E,R,L,$\cos\theta$) are
generated from the density functions $f_j(E,R,L,\cos\theta)$. The index $j$
stands for the 3 different contributions.

\begin{figure}[hbt]
  \centerline{\epsfig{figure=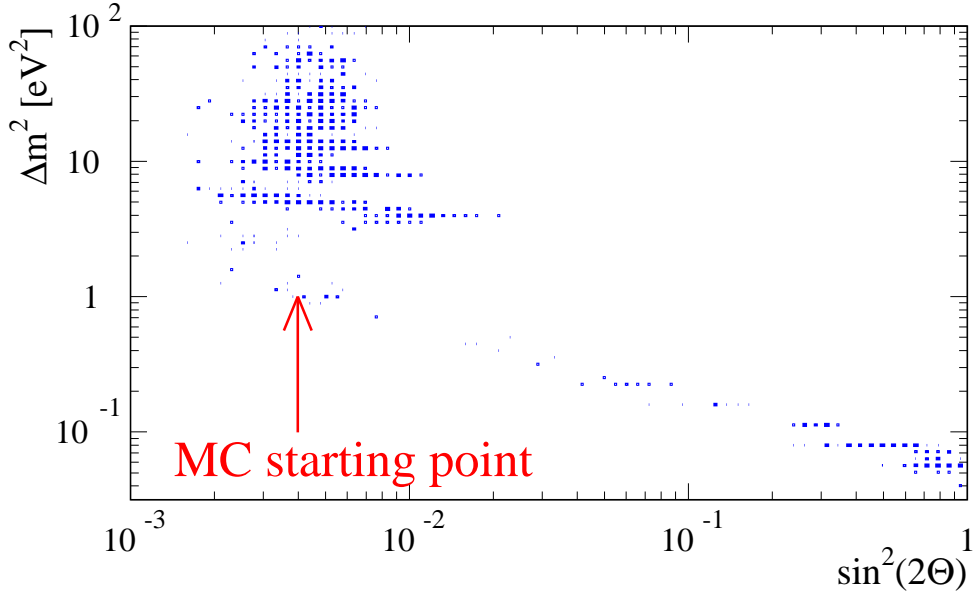,width=13.0cm}}
  \caption{Positions in (\sit,\Dm) of the maximum of the likelihood 
	functions for MC samples created with an \numubnueb\ oscillation 
	contribution based on $(\sit_H=4.2\cdot 10^{-3},\Dm_H=1eV^2)$.}
  \label{feld1}
  \end{figure}

After an event sample is generated, the sample is analysed
in exactly the same way as the experimental sample, i.e. the logarithm of
the likelihood function (\ref{evt-lhd}) is calculated as a function of
(\sit,\Dm). Figure~\ref{feld1} shows the distribution 
of the maxima $(\sit_m,\Dm_m)$ of 1000 MC generated samples with 
$(\sit_H,\Dm_H)=(4.2\cdot 10^{-3},\Dm=1eV^2)$.
The maxima are spread over a wide range in (\sit,\Dm) indicating already
the limited capability to determine a small area in (\sit,\Dm) on the
basis of the LSND event sample.

\begin{figure}[hbt]
  \centerline{\epsfig{figure=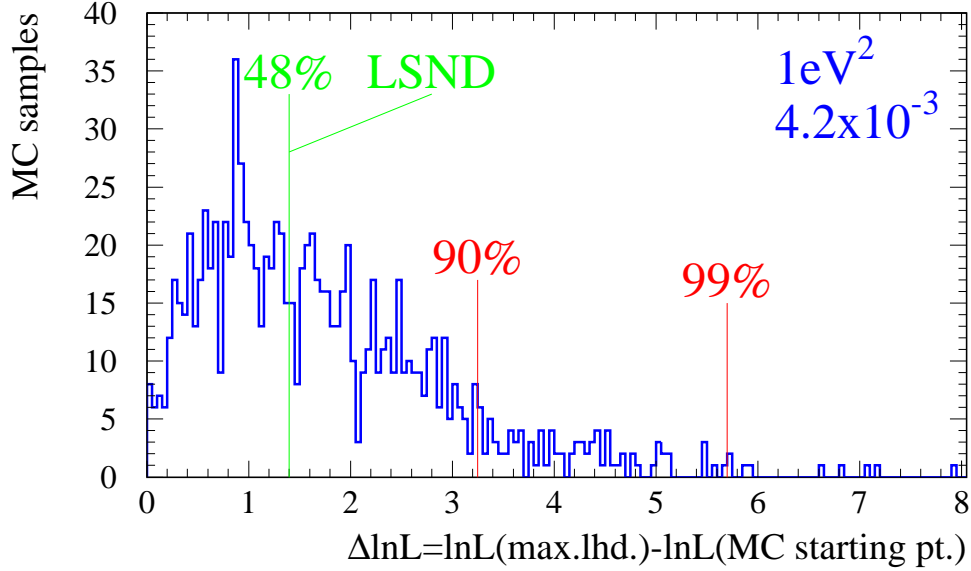,width=13.0cm}}
  \caption{Differences in $lnL$ between the actual maxima and the values
	at the MC starting point. Also indicated is the difference of
	the logarithmic likelihood function $\Delta lnL = 
	lnL(\sit_m,\Dm_m) - lnL(4.2\cdot 10^{-3},1eV^2) = 1.4$
	for the LSND sample.}
  \label{feld2}
  \end{figure}

To construct confidence regions, 
the distribution shown in figure~\ref{feld2} is central and should
be read in the following way: To include the oscillation hypothesis 
$(\sit_H,\Dm_H)$
with a probability (frequency of occurrence) of 90\%, 
the area in (\sit,\Dm) has to be defined by cutting $lnL$ at a value
of $\Delta lnL(90\%) = 3.25$ for each individual likelihood function. This
statistic as a function of $\Delta lnL$ shows the spreading of the
maximal value of $lnL$ compared to a given pair of oscillation parameters.
If, for a given experiment, the value $\Delta lnL^{exp}$ is smaller than
$\Delta lnL$ obtained for a specific hypothesis, such a parameter combination
$(\sit_H,\Dm_H)$ would be included in the region of 90\% confidence.
For the LSND logarithmic likelihood function, the difference is
$\Delta lnL = lnL(\sit_m,\Dm_m) - lnL(4.2\cdot 10^{-3},1eV^2) = 1.4$
As shown in figure~\ref{feld2}, in 52\% of all MC samples
$\Delta lnL$ is expected to be larger than 1.4 demonstrating that 
$(4.2\cdot 10^{-3},1eV^2)$ is clearly within the 90\% confidence region of the
LSND experimental result.

As $\Delta lnL(90\%)$ is itself a function of the parameters
$(\sit_H, \Dm_H)$, the generation of MC samples has to be repeated for all 
possible parameter combinations (\sit,\Dm) under consideration.
This task, however, is almost impossible due to the large computing time
necessary to construct the statistic in $\Delta lnL$ for each point.
The creation and likelihood evaluation of 1000 samples as described above
takes about 500 hours CPU time on the SGI origin 200 with R10000
processor available
within the LSND computer cluster. The strategy was therefore to construct
these statistics for representative pairs (\sit,\Dm) and interpolate
the obtained $\Delta lnL(90\%)$ values.

The normalized distribution in figure~\ref{feld2} is named $C'(\Delta lnL)$ 
and the variable
\begin{equation}
 lnL(\sit_m,\Dm_m) - lnL(\sit_H,\Dm_H) = \Delta lnL \equiv  \Delta \quad .
\end{equation}
Plotting the normalized integration of $C'$ as function of $\Delta$ defined as
\begin{equation}
	C(\Delta) = \frac{\int_0^{\Delta} C'(x)dx}
			 {\int_0^{\infty} C'(x)dx}
\end{equation}
allows an easy extraction of the 90\% confidence value
$\Delta^{90}$ for which $C(\Delta^{90}) = 0.9$.
  \begin{figure}[hbt]
  \centerline{\epsfig{figure=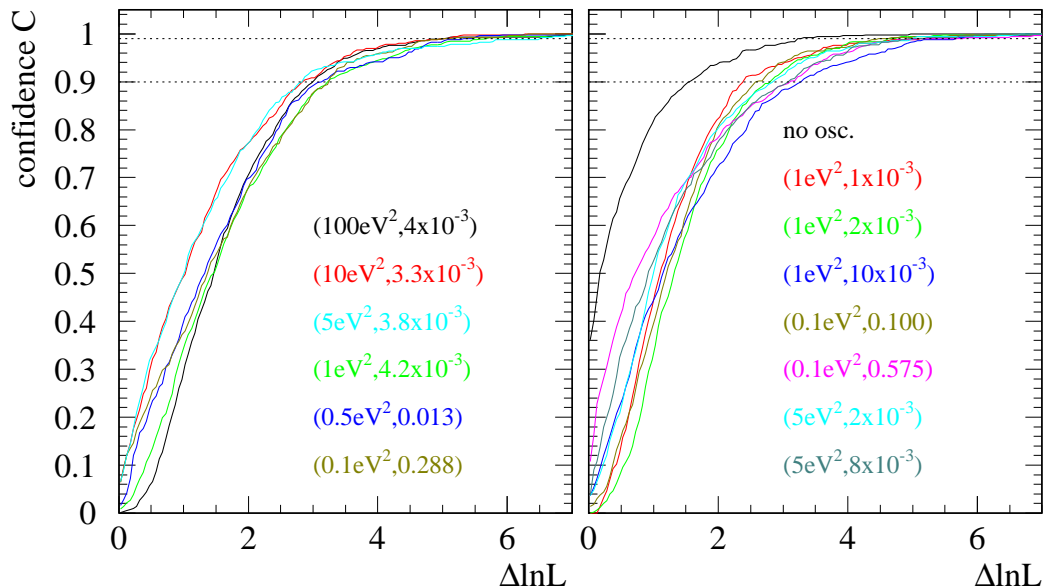,width=14.0cm}}
  \caption{Cumulative distributions $C(\Delta lnL_L)$ for various starting
	points $(\sit_H, \Dm_H)$. The left plot shows the distributions $C$
	for hypotheses with high likelihood for the LSND sample whereas
	the right figure is based on 'unlikely' starting hypotheses.
	The intersection of $C$ with the dotted lines can be used to extract 
	the $\Delta^{90}$ and $\Delta^{99}$ values.}
  \label{feld3}
  \end{figure}
Shown in figure~\ref{feld3} are some distributions $C(\Delta_L)$ including
the one for $(\sit_H=4.2\cdot 10^{-3},\Dm_H=1eV^2)$ for the LSND analysis. 
Note that these $C$ distributions could be quite different. 
There are two major conclusions to be drawn from figure~\ref{feld3}: 
It is obvious that the approximations of constant $\Delta^{90}=2.3$ and 
$\Delta^{99}=4.6$ under the assumption of a Gaussian likelihood function
with 2 independent parameters do not hold anymore\footnote{Note for further
comparison that $\Delta^{95}=3.0$ under this assumption.}. 
On the other hand, although there
are differences in the shape of $C(\Delta_L)$ depending on $(\sit_H, \Dm_H)$
the variations are not too large, even by comparing extreme hypotheses
like $(\sit_H=0.1, \Dm_H=0.1eV^2)$ or $(\sit_H=1\cdot10^{-2}, \Dm_H=1eV^2)$
with 25 or 170 oscillation events in average within their MC samples,
respectively. This fact is mainly due to the relative high statistics of
the event samples and their individual components which ensure relatively
stable results of the likelihood analysis.
 
\begin{figure}[hbt]
  \centerline{\epsfig{figure=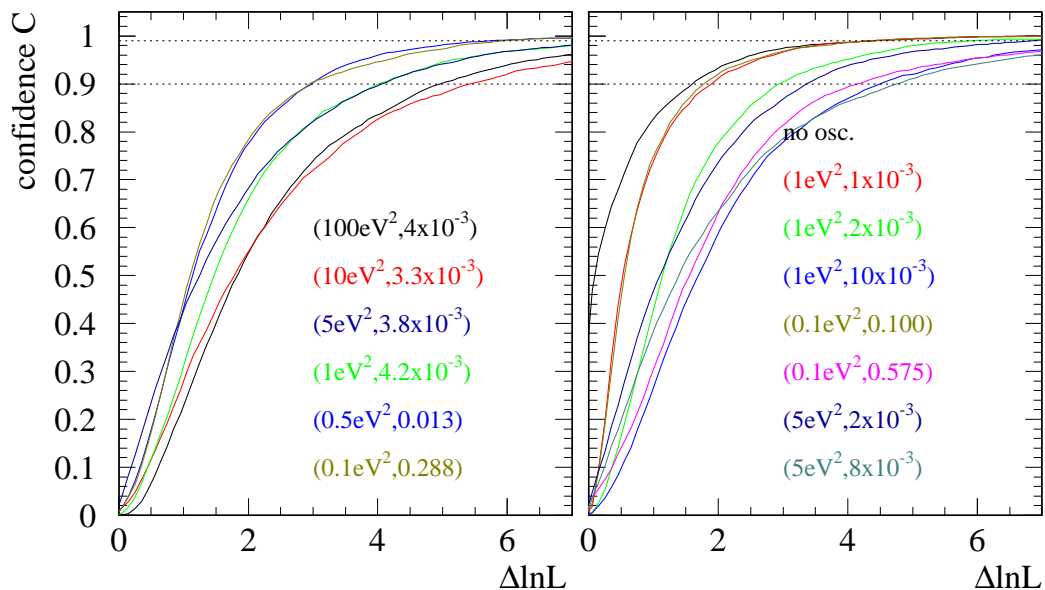,width=14.0cm}}
  \caption{Cumulative distributions $C(\Delta lnL_K)$ for the same starting
	points $(\sit_H, \Dm_H)$ as in figure~\ref{feld3}, but for the
	KARMEN experimental statistics and conditions. Note the larger
	spread of $\Delta^{90}$ and $\Delta^{99}$ values.}
  \label{feld3b}
  \end{figure}

The situation for KARMEN concerning the $C(\Delta_K)$ distribution is 
different for two reasons. KARMEN2 has seen 8 events so far. The
samples created by MC consequently also consist of this small statistic.
Therefore, the distributions $C$ and the values $\Delta^{90}_K$ as functions
of \sit\ and \Dm\ have larger variations. This is demonstrated in
figure~\ref{feld3b} for the same starting points as for LSND. 
On the other side, the small event sample reflecting the very low background 
of KARMEN2 allows, in reasonable computing time, to create a fine grid
of starting points $(\sit_H, \Dm_H)$ with a high number of MC samples per
grid point. This leads to an accurate two dimensional distribution of the
values $\Delta^{90}_K$ and $\Delta^{99}_K$.

On the basis of the distributions $C(\Delta)$ the values $\Delta^{CL}$
for a given confidence level CL are given for the calculated 
$(\sit_H, \Dm_H)$. The next step consists of interpolating from these
points to span a surface of $\Delta^{CL}(\sit,\Dm)$ for each experiment. 
For LSND, there are, due to the limited CPU time and the large samples, 
14 representative points with 1000 MC samples in the region of high 
likelihood, including the extreme of no oscillations. 
For KARMEN there are $90\times72$ points over the (\sit,\Dm) region of 
interest with 4000 MC samples created at each grid point. 
The corresponding confidence regions for both experiments were then obtained 
by cutting the logarithmic likelihood function $lnL(\sit,\Dm)$ at values of 
$\Delta^{CL}(\sit,\Dm)$ below the absolute maximum of $lnL$.

\begin{figure}[hbt]
  \centerline{\epsfig{figure=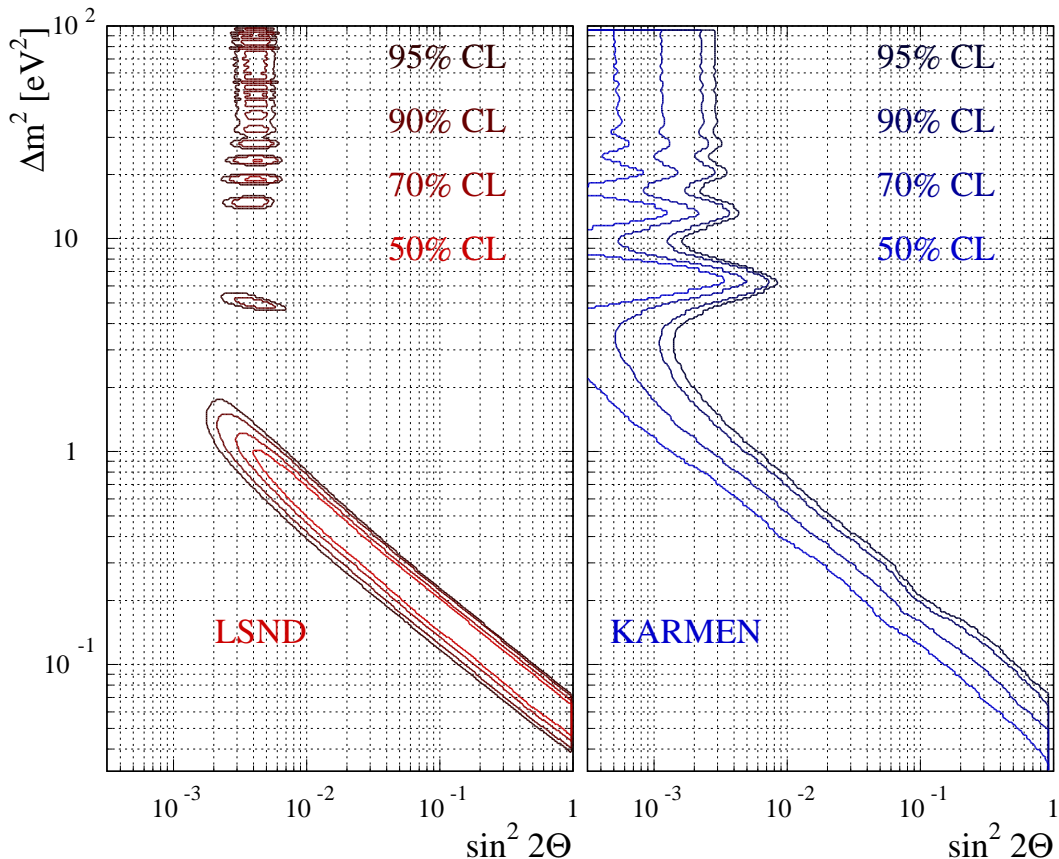,width=14.0cm}}
  \caption{Regions of some confidence levels in (\sit,\Dm) for the
	LSND and KARMEN experiment. For LSND, the innermost contour
	surrounds the region of 50\% confidence with increasing confidence
	level moving to the outer contours. For KARMEN, the rightmost
	line describes the 95\% confidence exclusion curve with lower 
	confidence going to the left. Areas to the right of the KARMEN curves 
	are excluded at the corresponding confidence level.}
  \label{feld4}
  \end{figure}

Figure~\ref{feld4} shows regions of 4 different confidence levels for
both experiments individually. For LSND, the innermost contour represents
the area of lowest confidence. For KARMEN the confidence level of excluded
areas increases with the curves from left to right. The lack of smoothness
of the lines reflects the limited number of grid points as well as the
statistics of samples generated per grid point $(\sit_H, \Dm_H)$.
This second limitation is also the reason why no \NNCL\ region is plotted.
Checking the distribution shown in figure~\ref{feld1} of 1000 MC samples
demonstrates that $\Delta^{99}_L$ for LSND is determined by the tail of
10 MC samples and has therefore a large uncertainty. The highest confidence
level deduced by these distributions which will be used in this context is
therefore 95\% confidence.

At \NCL , each individual experimental outcome was compared with other 
experiments.
  \begin{figure}[hbt]
  \centerline{\epsfig{figure=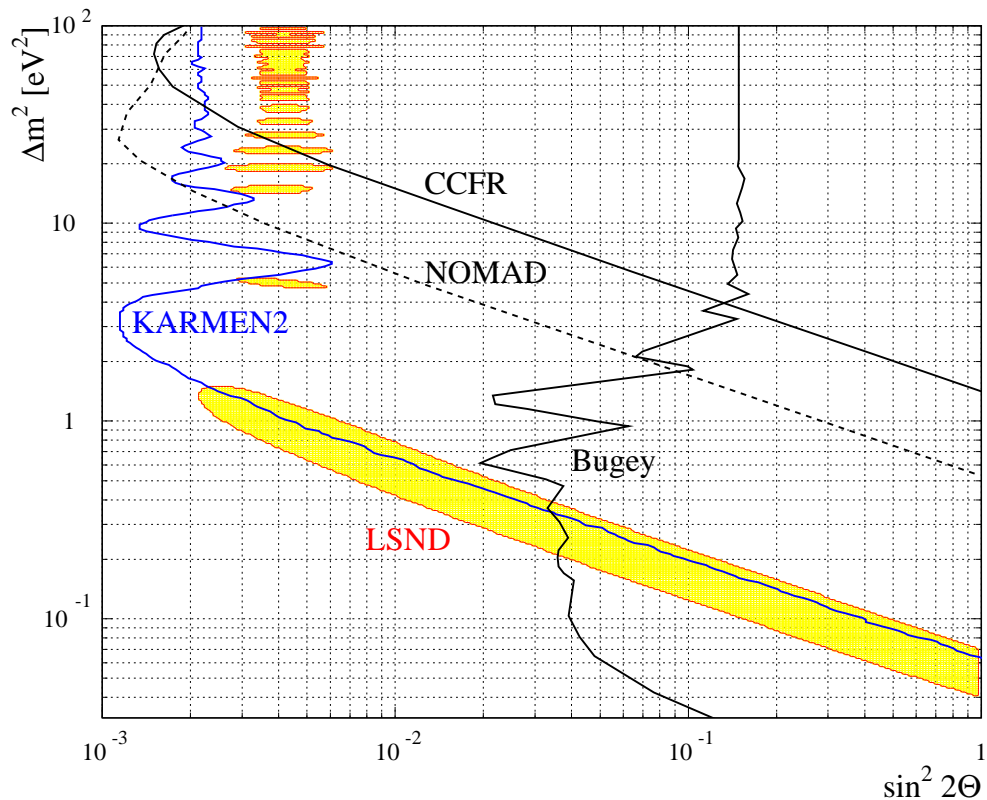,width=13.0cm}}
  \caption{LSND \NCL\ region in comparison with other \NCL\ exclusion
	curves in the corresponding (\sit,\Dm) area. The extraction of
	the \NCL\ curves for NOMAD, CCFR and Bugey are not based on the
	frequentist approach used for KARMEN and LSND.}
  \label{feld5}
  \end{figure}
Figure~\ref{feld5} shows the oscillation parameters inside the \NCL\
LSND region and the \NCL\ limits from KARMEN2 and other experiments.
Notice that the limits of the Bugey \nuebx\ search \cite{bugey}, the
CCFR combined \numunue\ and \numubnueb\ search \cite{ccfr} and the
preliminary results from the NOMAD \numunue\ search \cite{nomad} are not 
based on this unified frequentist approach by Feldman and Cousins.
Comparing the LSND \NCL\ region with the region defined by a constant
$\Delta lnL=2.3$ (see figure~\ref{bin-evt} for 1993--98 data), solutions with
high \Dm\ 'reappear'. However, due to the relative steepness of the logarithmic
likelihood function of the 1993--98 LSND evaluation, the changes from
$\Delta lnL=2.3$ to the correct $\Delta lnL(\sit,\Dm)$ do not lead to
a dramatically different confidence region in (\sit,\Dm).

One of the most misleading but nevertheless very frequently used 
interpretation of the LSND and KARMEN results
is to take the LSND region left of the KARMEN exclusion curve as 
area of (\sit,\Dm) 'left over'. Such an interpretation, though appealingly
straight forward, completely ignores the information of both likelihood
functions and reduces them to two discrete levels of individual 90\%
confidence. To be able to correctly combine the two experimental results
and extract the combined confidence regions,
we have to go some steps back to the original information of
the distributions $C'_K(\sit,\Dm)$ for KARMEN and $C'_L(\sit,\Dm)$ for LSND.
This is the task for section~\ref{methods}. 

\subsection{Combining likelihood functions}\label{combi_lnl}

\begin{figure}[hbt]
  \centerline{\epsfig{figure=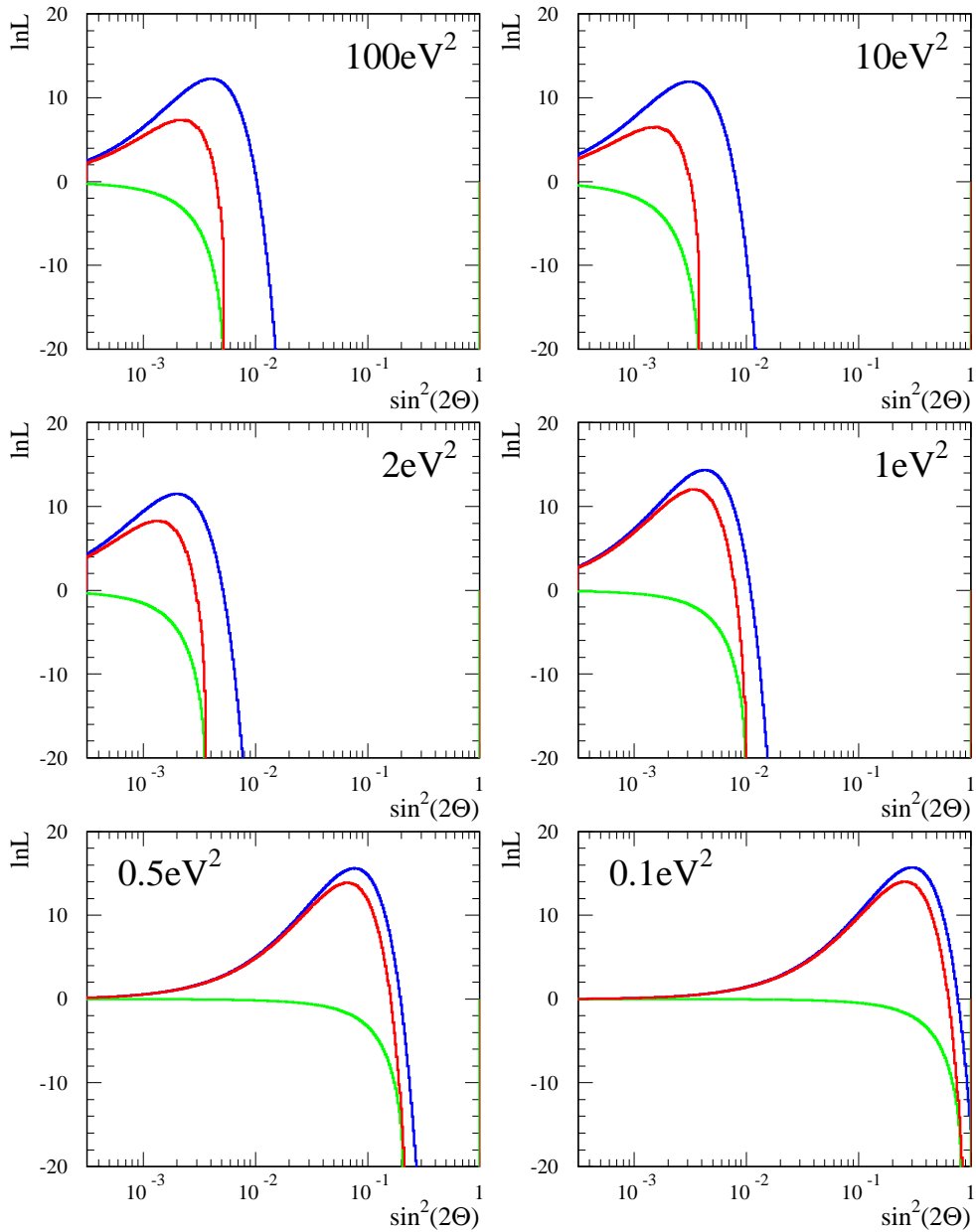,width=13.0cm}}
  \caption{Slices of constant \Dm\ of the logarithmic likelihood functions
	for KARMEN (leftmost or green), LSND (rightmost or blue) and
	the combination (middle or red). For definition of $lnL$ see text.}
  \label{comb_lhd}
  \end{figure}

It is a well known procedure to multiply the likelihood functions of two
independent experiments in order to combine the experimental results.
  \begin{figure}[hbt]
  \centerline{\epsfig{figure=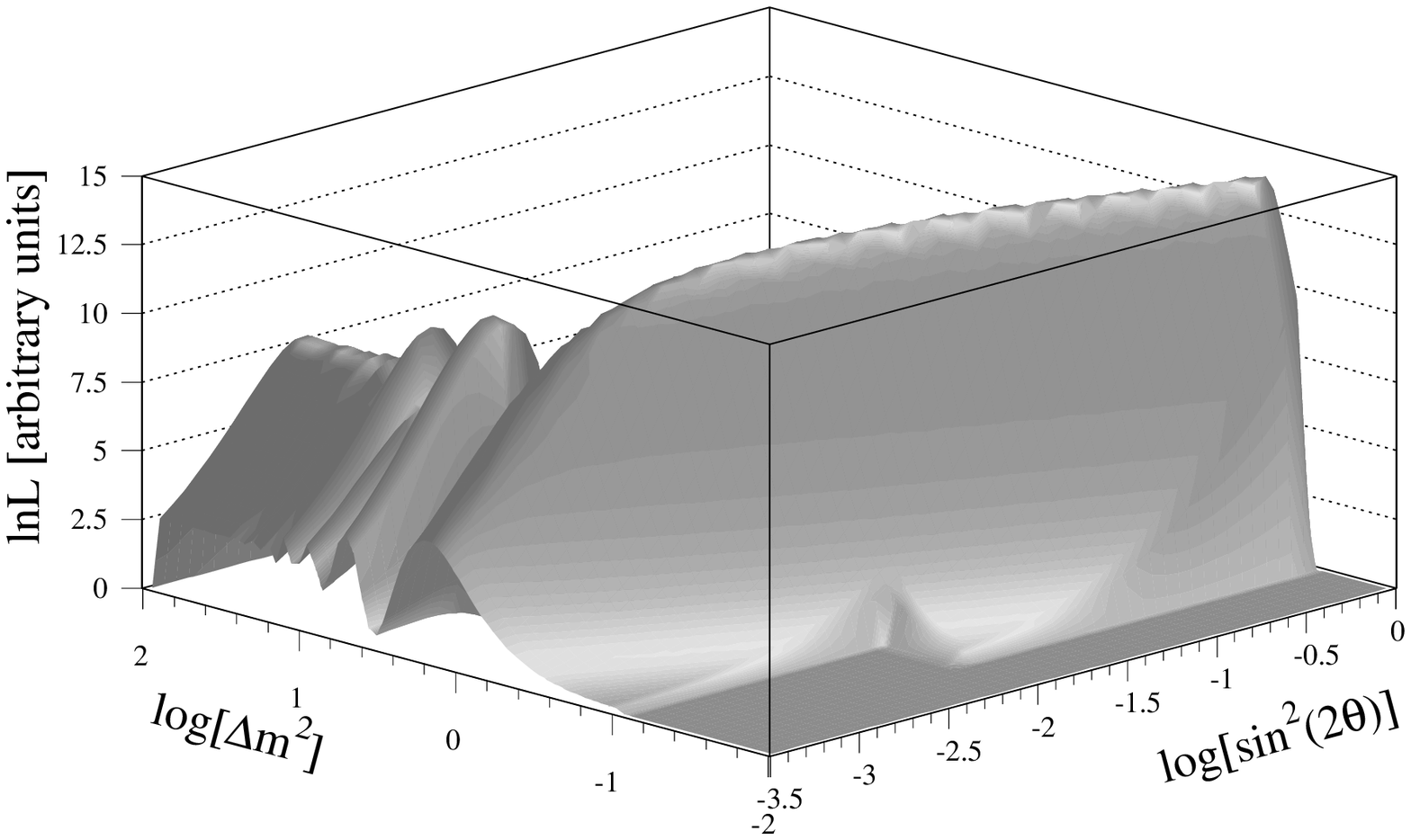,width=13.0cm}}
  \caption{Combined logarithmic likelihood function $lnL(\sit,\Dm)$
	as defined in equation~(\ref{kl_lhd}).}
  \label{kl-lhdfunction}
  \end{figure}
Instead of multiplying the likelihood functions, an equivalent way is to
add the logarithms. As already indicated in figures~\ref{k-lhdfunction} 
and \ref{l-lhdfunction}, there is some freedom in choosing the absolute
scale of $lnL$. A convenient presentation of $lnL$ is to normalize the 
individual functions $lnL_K$ and $lnL_L$ to a point in (\sit,\Dm) where
they are equally sensitive to a potential signal. In our case of the 
oscillation search this corresponds to values of $\sit=0$. A stringent 
exclusion would then lead to only negative values of $lnL$ whereas a strong
signal leads to a significant maximum with a positive value of $lnL$
\footnote{Note, however, that the absolute values of $lnL$ have no direct
meaning. Information can be obtained only by comparing the values within
the (\sit,\Dm) parameter space or with the individual experimental likelihood
functions.}.
Hence, the combined logarithmic likelihood function can be expressed as
\begin{eqnarray} \nonumber
 lnL(\sit,\Dm) &=& \{ lnL_K(\sit,\Dm)-lnL_K(\sit=0) \}  \\ &+&
		   \{ lnL_L(\sit,\Dm)-lnL_L(\sit=0) \}
\label{kl_lhd}
\end{eqnarray}
Figure~\ref{kl-lhdfunction} shows the combined function $lnL(\sit,\Dm)$
with a maximum of $lnL(\sit=0.78,\Dm=0.058)=14.1$ on a long flat 'ridge'
of low \Dm\ values. Figure~\ref{comb_lhd}
shows slices for some values of \Dm\ for the three functions 
$lnL_K(\sit,\Dm)-lnL_K(\sit=0)$ (leftmost or green curves), 
$lnL_L(\sit,\Dm)-lnL_L(\sit=0)$ (rightmost or blue curves) and $lnL$ as defined
in equation~\ref{kl_lhd}. The function $lnL(\sit,\Dm)$
allows a direct qualitative interpretation of the experiments:
There is a clear maximum of the combined likelihood function with a positive
value of $lnL$ favoring overall the evidence for oscillations given by
LSND. On the other hand, compared to the individual LSND maximum, 
$lnL_L$, the negative KARMEN result reduces the maximal value by 1.6 units 
(see figure~\ref{comb_lhd} for $\Dm=0.1$\,eV$^2$) which corresponds to a 
reduction to only 20\% 
of the original maximal likelihood. This reduction of the global maximum is a 
direct reflection of the general disagreement of the two experimental results.
From figure~\ref{comb_lhd} it is seen that
for low \Dm\ the position in \sit\ of the maximum is not substantially
shifted. In contrast, for larger \Dm\ the negative influence of the
KARMEN result clearly shifts the maximum in \sit\ and strongly reduces the LSND
likelihood value. It also increases the difference $\Delta lnL$ to the global 
maximum which is an important fact in terms of the statistics $C'(\Delta)$
and demonstrates that values of $\Dm>2$\,eV$^2$ have a much smaller likelihood 
than some combinations (\sit,\Dm) in the low \Dm\ region.

Although these observations help in assessing the combination of the two
experiments, probability statements cannot be deduced from the above
arguments. However, an evaluation of quantitative confidence regions can be 
based on the distributions $C'(\Delta)$, which is shown below.

\subsection{Methods to combine both experiments}\label{methods}

\begin{figure}[hbt]
  \centerline{\epsfig{figure=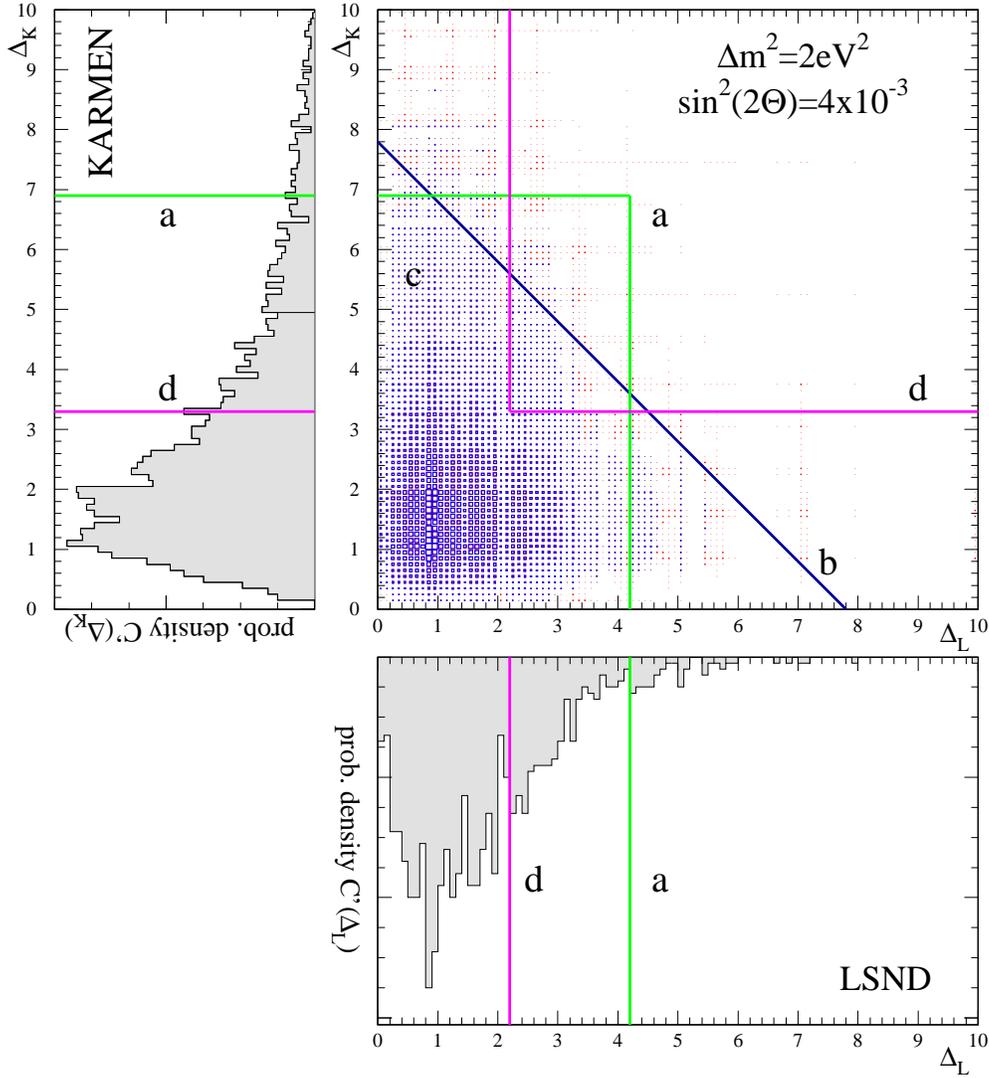,width=13.0cm}}
  \caption{Box plot of the two dimensional distribution 
	$C'(\Delta_L,\Delta_K)$ for a given
	oscillation parameter combination $(\sit=4\cdot 10^{-3},\Dm=2eV^2)$
	and its projections for the individual experiments. The different
	combining methods indicated (a) through (d) are described in the text.}
  \label{comfeld_t}
  \end{figure}

In this section we describe 4 different methods to extract areas
in (\sit,\Dm) of a certain confidence level CL. Though they
can be derived analytically we follow a more phenomenological approach.
The methods are based on different ways of ordering in a two dimensional 
space created by the individual
statistics of the two experiments, $C'_L$ and $C'_K$.
The assumption that the two experiments LSND and KARMEN 
are independent is well justified. Therefore,
a two dimensional distribution $C'(\Delta_L,\Delta_K)$ can be constructed
from the one dimensional normalized distributions $C'(\Delta_L)$ and
$C'(\Delta_K)$ by an inverse projection. A box plot of $C'(\Delta_L,\Delta_K)$
and its original functions $C'$ are shown in figure~\ref{comfeld_t}
for an example of a chosen parameter combination of 
$(\sit=4\cdot 10^{-3},\Dm=2eV^2)$. The different lines in
figure~\ref{comfeld_t} correspond to the limits for \NCL\ of the 
different methods described below. 

\subsubsection{Method a} 

This method combines LSND and KARMEN by integrating the
distributions for both experiments $i=K,L$ individually:
\begin{equation}
  \int_0^{\Delta_i^{CL}} C'(\Delta_i)d\Delta_i = CL \qquad i=K,L
\end{equation}
This corresponds to a rectangle in $(\Delta_L,\Delta_K)$ defined
by the side lengths $\Delta_K^{CL}$ and $\Delta_L^{CL}$. The
combined confidence is then $CL_{comb}=(CL)^2$. To obtain a confidence level
of $CL_{comb}=0.9$ we therefore have to determine $\Delta_i^{95}$.
The lines in figure~\ref{comfeld_t} labeled (a) show these values
$\Delta_i^{95}$ and the resulting rectangle in $(\Delta_L,\Delta_K)$.
If the experimental value $(\Delta_L^{exp},\Delta_K^{exp})$ lies within this
rectangle the parameter combination $(4\cdot 10^{-3},2eV^2)$ is included
in the combined \NCL\ region. This method can be expressed also by taking
the overlap of the $\sqrt{CL}$ confidence regions of both experiments
to deduce the combined $CL$ confidence region.

\subsubsection{Method b}

The second method is based on the combined statistic $C'(\Delta)$ with 
$\Delta=\Delta_L+\Delta_K$ defined as the convolution of the individual ones
\begin{equation}
  C'(\Delta) = \int_0^{\Delta} C'_L(\Delta_L) \cdot
	C'_K(\Delta - \Delta_L) d\Delta_L \quad .
\end{equation}
The confidence value $\Delta^{CL}$ is then defined by integration of $C'$:
\begin{equation}
  \int_0^{\Delta^{CL}} C'(\Delta)d\Delta = CL \quad .
\end{equation}
For a given $CL$, the limit corresponds to a diagonal line in 
figure~\ref{comfeld_t}, where (b) indicates $\Delta^{90}$ for this
specific (\sit,\Dm). The value $\Delta^{exp}=\Delta_L^{exp}+\Delta_K^{exp}$ 
is then compared with this $\Delta^{90}$. If $\Delta^{exp}\le \Delta^{90}$
the combination $(4\cdot 10^{-3},2eV^2)$ is accepted at a 90\% confidence
level. Such an approach in $(\Delta_L,\Delta_K)$ corresponds to an
ordering along lines of constant combined likelihood, $\Delta$ below the
two maxima of the likelihood functions.

\subsubsection{Method c} 

This method is based on an ordering principle of the elements 
$C'(\Delta_L,\Delta_K)$, i.e. the frequency or probability of occurrence of 
$(\Delta_L,\Delta_K)$. This differs to integrating starting at $\Delta=0$
as it is done in the previously described approaches. For a given confidence
level $CL$, combinations $(\Delta_L,\Delta_K)$ are added up in descending
order starting with
the highest probability of occurrence $C'$ until a fraction of $CL$
of the total $\int C'(\Delta_L,\Delta_K)d\Delta_L d\Delta_K$ is reached. In 
figure~\ref{comfeld_t} this subset $S$ of all $(\Delta_L,\Delta_K)$ is shown 
in blue. If $(\Delta_L^{exp},\Delta_K^{exp}) \in S$, the combination 
(\sit,\Dm) under consideration is included in the confidence region.

\subsubsection{Method d} 

The last method is in its ansatz not principally different to method b,
but results in a confidence region dramatically different to those
obtained by all other methods. Instead of taking the overlap of two regions
of $\sqrt{CL}$ confidence, the individual regions of $1-(1-CL)^2$ 
confidence are added to form the
combined region of $CL$ confidence. For a \NCL\ this means adding 
(mathematically building the .OR. of) the regions of 68.4\% 
individual confidence. In a graphical view, this is demonstrated by the 
line labelled (d) in figure~\ref{comfeld_t}. 

\subsubsection{Discussion} 

It is instructive to discuss the differences of the methods by
comparing the corresponding areas of the $(\Delta_L,\Delta_K)$ plane
(see figure~\ref{comfeld_t}) by each method. The triangle defined by
(b) and the rectangle defined by (a) have almost the same area. In their
corners with high values of $\Delta_i$ they allow experimental outcomes
which are very unlikely, at least for one experiment. This drawback
is overcome by the method (c) of ordering along probability of occurrence
which has the disadvantage of principally
disfavoring the unlikely, but very best fits of very small $\Delta_i$.
On the other side, the convolution method integrates along contours of 
constant likelihood for the combined likelihood function which is a very 
plausible procedure. The easiest and most straight forward method may
be method (a), and as we will see, leads to confidence regions very similar
to those obtained by the convolution or ordering method.

In section~\ref{combiregio} we will show the resulting confidence regions
in (\sit,\Dm) for all these methods and provide further discussion and
interpretation of the approaches.

\subsection{Combined regions of confidence}\label{combiregio}

\begin{figure}
  % old figure with LSND 2.3/4.6 areas: comfeld_vier.eps
  \centerline{\epsfig{figure=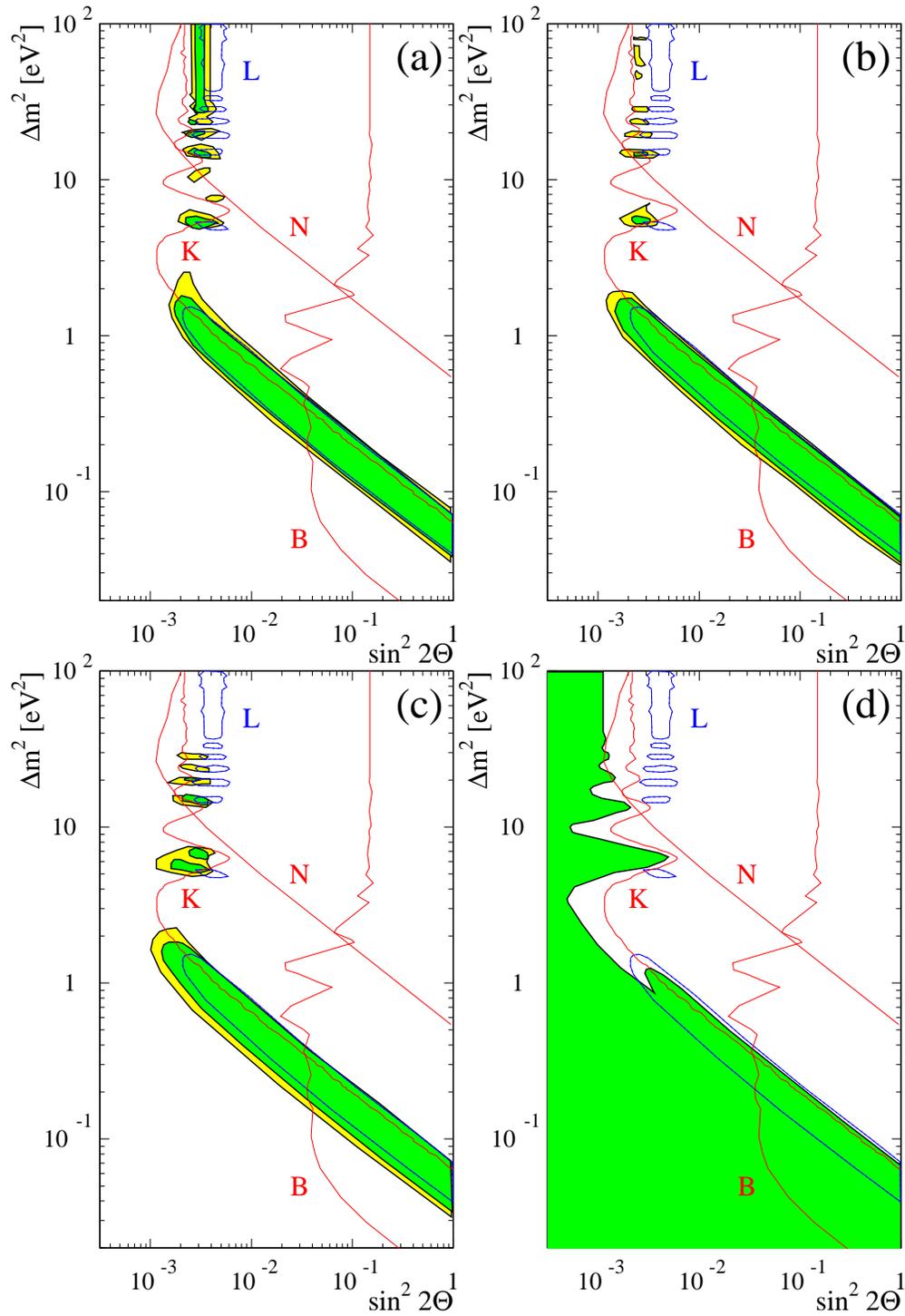,width=13.0cm}}
  \caption{Regions of 90\% and 95\% confidence for KARMEN and LSND
	combined as well as individual results of different experiments.
	See text for further explanations.}
  \label{comfeld_v}
  \end{figure}

The combined regions of 90\% and 95\% confidence are shown in 
figure~\ref{comfeld_v} as green and yellow areas in (\sit,\Dm). The figures
(a) through (d) correspond to the methods (a) through (d) described in section
\ref{methods}. Also shown for comparison are the individual experimental
results: The KARMEN \NCL\ exclusion curve (K) and the LSND \NCL\ region (L)
according to the frequentist approach (see figure~\ref{feld5})
as well as the exclusion curves of the two experiments Bugey \nuebx\ (B) and 
NOMAD \numunue\ (N).

Comparing the results of methods (a) to (c) (in short called overlap, 
convolution and ordering), the confidence regions have only minor
differences. High \Dm\ solutions are not excluded at 95\% 
confidence, although the convolution and ordering methods clearly favor 
$\Dm < 10$\,eV$^2$. The confidence
region for $\Dm < 2$\,eV$^2$ is almost identical for all combinations. At 
first sight, these regions are even similar to the \NCL\ region of LSND only
(see lines indicated with L in figure~\ref{comfeld_v}), 
however the combined \NCL\ region extends to smaller values of \sit\ in the 
low \Dm\ region. For large \Dm , the combined region is reduced 
and shifted to smaller mixing values. Although there are regions at 
$\Dm >2$\,eV$^2$ within a \NCL\ these solutions have considerably smaller
likelihood than along the 'ridge' at low \Dm , as was discussed in 
section~\ref{combi_lnl}. This argument is 
underlined if regarding regions of combined confidence at an 80\% 
confidence level. At such a level, none of the methods (a) through (c)
include solutions above $\Dm=2$\,eV$^2$.

As mentioned before, the limited statistics of the LSND frequentist analysis
creating the distributions $C'$ does not allow to calculate \NNCL\ regions
with the needed accuracy. However, the only minor extension of the \NCL\
area by the extracted combined 95\% confidence region indicates the
statistical significance of the positive LSND result. An oscillation
scenario according to this combined statistical analysis is clearly
compatible with both experiments. Of course, this does not account for a 
potential systematic error or misinterpretation of the beam excess seen 
in LSND.

Figure~\ref{comfeld_v}(d) shows a very distinct region of 90\% confidence.
It is a very controversial way of combining two experiments if their central
statements are different. By selecting relative high confidence (70\%) 
regions of each experiment and building a common area by adding these regions,
the underlying interpretation tends to choose one experiment over
the other. As discussed in section~\ref{methods} for the other methods,
this approach allows combinations of $(\Delta_L,\Delta_K)$ where one value
is not restricted at all, i.e. extremely unlikely points in the
constructed two dimensional statistics space of $C'(\Delta_L,\Delta_K)$.
This method will, with better individual experimental statistics,
ultimately lead to two distinct areas forming a combined \NCL\ which does
not help in making a decision about the statistical compatibility of
two experiments. Although this method results in a correct coverage,
it is therefore obviously disfavored. 

\section{Conclusion and outlook}\label{conclu}
% --------------------------------------------

The data sets of both the LSND and KARMEN experiment were analysed with
a maximum likelihood method. The definition of the LSND likelihood function
was changed from a combined likelihood of bin contents to a product of
event based likelihoods allowing the backgrounds to float according to
their expectation value and its uncertainty. This improvement led to slightly
lower values of \sit\ for a given \Dm .

For the first time, a frequentist approach based on \cite{cous} was
applied to determine confidence regions of correct coverage for the
LSND experiment. It is shown that in the case of a likelihood function
depending on the oscillation parameters \sit\ and \Dm , the
approach assuming a two dimensional Gaussian likelihood function
is only a rough approximation and does not lead to correct coverage.

As both the KARMEN and LSND experimental data were analysed with a likelihood 
function and
the statistics to deduce confidence regions were built in the same manner,
it is possible to combine the likelihood functions and extract combined
confidence regions based on a combination of the individual statistics
created by Monte Carlo procedures. These regions are regions of correct 
coverage in terms of a frequentist approach.

The combined confidence regions in (\sit,\Dm) extracted from different methods
to combine the experiments are very similar (if we exclude the controversial
and not really convincing method (d)). Though graphically not very different 
from what one would expect in a naive approach using the individual \NCL\
regions, a more detailed look shows that there are subtle differences.
In figure~\ref{feld5} there are no LSND areas left to the KARMEN exclusion
curve for $\Dm>2$\,eV$^2$. The combined \NCL\ regions in 
figure~\ref{comfeld_v}(a-c) however include some areas of larger \Dm\ although 
the likelihood value $lnL$ is much smaller than for $\Dm<2$\,eV$^2$ therefore
favoring the low \Dm\ solutions. Requiring a more stringent confidence level 
of 80\% or less, only solutions with $\Dm<2$\,eV$^2$ remain.

The results of this analysis remain preliminary, as stated in the introduction.
This is not due to the statistical analysis itself but to the data sets
used. However, the work described here is the first statistical
analysis combining both the LSND and KARMEN experimental outcomes and
shows the feasability and results of such a method. As there are other
experiments like NOMAD, CCFR and Bugey sensitive in part to the confidence 
region in (\sit,\Dm), a complete analysis should also include these results
on the basis of the same statistical analysis. This implies, however, the
detailed knowledge of experimental data of these experiments not accessible
to the author. In addition, the exclusion curve from the Bugey experiment is
based on the disappearance search \nuebx . Combining this experiment 
correctly with the appearance results of \numubnueb\ or \numunue\ in terms
of mixing angles would therefore also require a full three or four dimensional 
(with a sterile neutrino) mixing scheme.

KARMEN2 has no evidence for oscillations and sets the most stringent 
experimental limits so far on the mixing 
\sit\ for a range of $0.2 \le \Dm \le 20$\,eV$^2$. However, the sensitivity 
of its result is not sufficient to completely cover the parameter region 
given by the LSND signal. 
It was shown quantitatively that there are areas of confidence 
in (\sit,\Dm) which are compatible with both experiments, especially
for $\Dm \le 2$\,eV$^2$. Such a statistical analysis is necessary 
to assess the two experimental outcomes in terms of (\sit,\Dm), 
however, further experimental investigations are needed.
The Booster Neutrino Experiment BooNE~\cite{boone} at Fermilab
will be built to check this controversial region of the oscillation 
parameters with high statistics and different systematics. Another experiment
is proposed at the CERN proton synchrotron~\cite{i216}. These experiments
may then resolve the issue of the KARMEN and LSND results
on \numubnueb\ oscillations.

\ack
% --

Many people have substantially contributed to this work.
First of all, I want to thank both collaborations in general for allowing
the use of the full experimental data for this analysis. In particular,
Markus Steidl provided me with the most recent data analyses of KARMEN2.
In numerous discussions the track for the analysis was set up.
Especially helpful were discussions and clarifications with Geoffrey B. Mills
and Steven J. Yellin's proposals on how to combine the individual statistics,
and I gratefully acknowledge their help. 

Last but not least I want to thank the Alexander von Humboldt-Foundation who, 
by granting me a Feodor-Lynen fellowship to visit Gerald T. Garvey, made 
possible a very stimulating and fruitful stay with my colleagues in Los Alamos.

\end{document}